\documentstyle[12pt]{article}
\def\le{\left}
\def\ri{\right}

\def\part{\partial}

\def\bq{\begin{eqnarray*}}
\def\eq{\end{eqnarray*}}
\def\beq{\begin{eqnarray}}
\def\eeq{\end{eqnarray}}

\def\ba{\begin{array}}
\def\ea{\end{array}}

\def\bil{\Bigl}

\def\ov{\overline}

\def\ep{\epsilon}

\def\cd{\cdot}

\begin{document}
\[ \]
\[ \]
\begin{center}
{\large \bf Formation of cosmological mass condensation within a FRW \\[0.1cm]
universe: exact general relativistic solutions}
\[ \]
{\bf S. Khakshournia\footnote{Email Address:
skhakshour@seai.neda.net.ir}}

Department of Physics, Sharif University of Technology, Tehran,
Iran

and

Nuclear Research Center, Atomic Energy Organization of Iran,
Tehran, Iran
\[\]
{\bf R. Mansouri\footnote{Email Address: mansouri@sharif.edu}}\\

Department of Physics, Sharif University of Technology, Tehran,
Iran

and

Institute for Studies in Physics and Mathematics, P.O. Box 5531,
 Tehran, Iran
 \[\]
\end{center}
\[ \]
\begin{center}
{\bf Abstract}
\end{center}
\hspace*{0.5cm}Within the framework of an exact general
relativistic formulation of gluing manifolds, we consider the
problem of matching an inhomogeneous overdense region to a
Friedmann-Robertson-Walker background universe in the general
spherical symmetric case of pressure-free models. It is shown
that, in general, the matching is only possible through a thin
shell, a fact ignored in the literature. In addition to this, in
subhorizon cases where the matching is possible, an intermediate
underdense region will necessarily arise.

\newpage
Large scale structure formation is a challenging field from both
theoretical and observational points of view. Motivated by the
observational data on the existence of large scale voids in the
universe, Sato and co-workers considered in a series of papers an
underdense spherical region immersed in a
Friedmann-Robertson-Walker (FRW) background universe applying the
thin shell formalism of general relativity[1]. On the other hand,
the appearance of holes with densities less than the average
cosmological density around rich spherical clusters of galaxies
has been seen by some authors [2-4]. In this paper, we model such
inhomogeneous large scale structures as a Lemaitre-Tolman-Bondi
(LTB) manifold with different density profiles glued to a
homogeneous pressure-free FRW background from which a sphere of
dust matter is removed. Our calculation is based on an exact
general relativistic formulation of gluing manifolds. We may
therefore consider our contribution as a generalization of the
work done by Olson and Silk[3]. Their calculation was made
basically within Newtonian dynamics
without being cautious about the matching conditions.\\
Consider first a spherical inhomogeneous region containing dust
matter, represented by a LTB cosmological model embedded in a
pressure-free FRW background universe with the uniform density
$\rho_{b}$. We choose the LTB metric to be written in the
synchronized comoving coordinates in the form[5]
\begin{eqnarray}
ds^{2}=-dt^{2}+\frac{R'^{2}}{1+E(r)}dr^{2}+R^{2}(r,t)(d\theta^{2}+\sin^{2}
\theta d\phi^{2}).
\end{eqnarray}
 The overdot and prime denote partial differentiation with respect to $t$
 and $r$, respect-\\ively, and $E(r)$ is an arbitrary real function such that
 $E(r)>-1$. Then the corresponding Einstein field equations turn out to be
\begin{eqnarray}
&&\dot{R}^{2}(r,t)=E(r)+\frac{2M(r)}{R} ,\\
&&\hspace*{0.6cm}4\pi\rho(r,t)=\frac{M'(r)}{R^{2}R'}.
\end{eqnarray}
 To be as general as possible, we let the density $\rho (r,t)$
to be a general function of $r$, i.e., it is not necessarily a
monotonic function of radial distance from the center of the
region with mass condensation. $M(r)$ is defined by
\begin{eqnarray}
M(r)=4\pi \int^{R(r,t)}_{0}\rho(r,t)R^{2}dR \cdot
\end{eqnarray}
Furthermore, in order to avoid shell crossing of dust matter during their
radial motion, we must have $R'(r,t)>0$. The solution to the above equations shows
that an overdense spherical inhomogeneity with $E(r)<0$ within $R$ evolves just like
a closed universe, namely it reaches to a maximum radius at a certain time, then the
expansion ceases and undergoes a gravitational collapse so that a bound object forms
in such a way.\\
Now, in the literature on large scale structures of the universe
it is always tacitly assumed that the junction of an overdense
region to a background FRW universe is a continuous one and there
is no need of a singular hypersurface along which the gluing is
made[6]. Let us denote by $\Sigma$ the (2+1)-dimensional timelike
boundary of the two distinct spherically symmetric regions glued
together. We will show that this is in general not true and in the
most cases of interest $\Sigma$ cannot be just a boundary surface
 but is a singular hypersurface carrying energy momentum. \\
We now write down the appropriate junction equation on $\Sigma$
expressing the jump of the angular component of extrinsic
curvature tensor $K^{\theta}_{\theta}$ across $\Sigma$ as [7,8]
\begin{eqnarray}
\epsilon_{in}\sqrt{1+\left( \frac{dR}{d\tau}\right)^{2}-
\frac{8\pi\overline{\rho}}{3}R^{2}}-
\epsilon_{out}\sqrt{1+\left( \frac{dR}{d\tau}\right)^{2}
-\frac{8\pi\rho_{b}}{3}R^{2}}\stackrel{\Sigma}{=}4\pi\sigma R \ ,
\end{eqnarray}
where $\stackrel{\Sigma}{=}$ means that both sides of the equality are
evaluated on $\Sigma$, $\tau$ is the proper time of the comoving observer on $\Sigma$, and
$\sigma$ is the surface energy density of the shell $\Sigma$. The sign functions are
fixed according to the convention $\epsilon_{in}(\epsilon_{out})= +1$ for $R$ increasing
in the outward normal direction to $\Sigma$, while $\epsilon_{in}(\epsilon_{out})= -1$ for
decreasing $R$. An average density for the inhomogeneous region is also defined as
\begin{eqnarray}
\overline{\rho} = \frac{M(r)}{{4\pi\over 3}R^{3}} \bil|_{\Sigma}.
\end{eqnarray}
Note that in general the matching is only possible if a thin layer is formed on the
boundary of the two manifolds, i.e. $\sigma \neq 0$, except for the case where
$\overline{\rho} = \rho_b$, and $\epsilon_{in} = \epsilon_{out}$.\\
Now, similar to the approach used by Berezin et al [8], we solve
Eq. (5) for $\le( \frac{dR}{d\tau}\ri)^{2}$ to obtain
\begin{eqnarray}
\left( \frac{dR}{d\tau}\right)^{2}=\left(
4\pi^{2}\sigma^{2}(\xi -1)^{2}+\frac{8\pi}{3}\rho_{b}\right) R^{2}-1,
\end{eqnarray}
where $\xi$ is defined by
\begin{eqnarray}
\xi\equiv\frac{\rho_{b}-\overline{\rho}}{6\pi\sigma^{2}} \cd
\end{eqnarray}
Noting that the surface energy density $\sigma$ on $\Sigma$ is
positive, we substitute Eq. (7) back into Eq. (5) to get
\begin{eqnarray}
\epsilon_{in}|\xi+1|-\epsilon_{out}|\xi-1|=2 \cdot
\end{eqnarray}
It is easily seen that the sign functions $\epsilon_{in/out}$ are determined
by the value of $\xi$:
\begin{eqnarray}
(\epsilon_{in},\epsilon_{out})=\left\{ \begin{array}{llll}
(+1,+1)&\mbox{for}&\xi>1,\\
(+1,-1)&\mbox{for}&|\xi|<1,\\
(-1,-1)&\mbox{for}&\xi<-1\cd \end{array} \right.
\end{eqnarray}
Now, $\epsilon_{out}$ may be related to different parameters of FRW universe. It is known
that[9]
\begin{eqnarray}
\epsilon_{out}=sgn \left( \frac{dr_{b}}{d\chi_{b}}+v_{b}H_{b}R\right) ,
\end{eqnarray}
where $H_{b}$ is the Hubble parameter of the background, $v_{b}$ is the
peculiar velocity of $\Sigma$ relative to the background. The two coordinates
$r_{b}$ and $\chi_{b}$ are related to each other as
\begin{eqnarray}
r_{b}(\chi_{b})=\left\{ \begin{array}{ll}\sin \chi_{b}&(k=+1,
\hspace*{0.55cm}\mbox{closed universe}),\\
\chi_{b}&(k=0,\hspace*{0.55cm}  \mbox{flat universe}),\\
\sinh\chi_{b}&(k=-1, \hspace*{0.55cm}\mbox{open universe})\cdot
\end{array}\right.
\end{eqnarray}
Depending on the density distribution within the shell,the
following cases may now be distinguished:\\
(I) Junction of an overdense region to an background FRW universe,
i.e., $\rho (r, \tau) > \rho _b$. Obviously $\xi < 0$. Therefore,
according to Eq. (10), we must have $\epsilon_{out} = -1$.
This leads to the following cases:\\
(1) $k = 0, -1; R < H^{-1}_{b}$. In both these cases it is easily
seen from Eqs. (11),(12) that $\epsilon_{out} = +1$. Therefore,
there is no matching of an overdense region smaller than the
Hubble radius to a background FRW universe with $ k = 0, -1$.\\
(2) $ k = 0, -1; R > H^{-1}_{b}$, and the peculiar velocity $v_{b}
< 0$. It is seen from Eq. (11) that $\epsilon_{out} = -1$[9].
Therefore, overdense structures with a radius greater than the
Hubble radius and $v_{b} < 0$ may be glued to a background FRW
with $k = 0, -1$. It is trivially seen that there is no
matching for the case of $v_{b} >0.$\\
(3)  $k = +1$. Taking the definition Eq. (12), we see from (11)
that it is possible to have $\epsilon_{out} = -1$. Therefore the
matching is possible without any restriction.\\
Except from this last case of a closed FRW background universe and
the case of a superhorizon overdense region discussed in (2), we
have seen that it is not possible to match an overdense spherical
region to a background FRW universe. Now, to look for more
possible matching let us assume the overdense structure to be
surrounded by an underdense region. Three other
cases may be distinguished now.\\
(II) $\overline \rho < \rho_{b}; k = 0, -1$. We have therefore an
overdense inhomogeneous region surrounded continuously by an
underdense region $\rho(r,t)<\rho_{b}$, i.e., a void described by
the same LTB metric. Now, note that the junction of the LTB region
at the underdense  void part of it to the background FRW is also
described by Eq. (5), while $\overline{\rho}$ is the overall
average density related to the total mass enclosed within the
boundary separating the hole
region of physical radius $R$ from the uniform background. \\
(1)  $R < H_{b}^{-1}$. It is seen from Eq. (11) that
$\epsilon_{out} = +1$, irrespective of the sign of the peculiar
velocity. Therefore, according to Eq. (10), the matching is
possible with $\xi > 1$ via a thin shell having $\sigma \neq 0$.\\
(2)  $R > H_{b}^{-1}$. In this case for the negative peculiar
velocity we have $\epsilon_{out} = -1$. Therefore the matching is
possible with $0 < \xi <1$, i.e., $\sigma \neq 0$. If the peculiar
velocity is positive we are faced with a case exactly the same as (1).\\
(III) $\overline \rho = \rho_{b}; k = 0, -1$. Now, the void
completely compensates the overdense region, so that the overall
mean density is equal to the background density.\\
(1)  $R < H_{b}^{-1}$. It turns out that $\sigma = 0$, i.e., the
matching is possible without formation of a thin shell. This
situation reminds us of the Einstein-Strauss type model [10] where
the overdense region is surrounded by the vacuum shell to
compensate the mass excess.\\
(2)  $R > H_{b}^{-1}$. For the negative peculiar velocity, we have
again $\epsilon_{out} = -1$. Both junctions with $\sigma =0$ and
$\neq 0$ are possible, as it is easily seen from Eq. (5). The case
$\epsilon_{in} = -1$ corresponds to $\sigma =0$ meaning no thin
shell, and the case $\epsilon_{in} = +1$ is only possible through
a thin shell. The case of positive peculiar velocity is similar to the (1) case.\\
(IV) $\overline \rho > \rho_{b}; k = 0, -1$. This is similar to the matching
where no void is in between, as discussed in the\\
cases considered in (I).\\
Table I summarizes the above results.\\

\begin{center}
{\small Table I. Classification of junctions of LTB spherical
inhomogeneity to the flat and open FRW background. An extensive
version of the paper with the detailed calculation and critical
review of the relevant literature is going to be published in
[11]}.
\end{center}
\[\ba{|c|c|c|} \hline &\ba{c} R>H_{b}^{-1}\mbox{,}\   v_{p}<0\\ (\ep_{out}=-1) \ea&
\ba{c} R<H_{b}^{-1}\ \mbox{or}\\R>H_{b}^{-1}\mbox{,}\ v_{p}>0\\ (\ep_{out}=+1) \ea\\
\hline
\ov{\rho}>\rho_{b}&\ba{c} \mbox{Thin shell}\\
\xi <1 \ea& \mbox{Junction impossible} \\ \hline
\ov{\rho}=\rho_{b}& \ba{l} \mbox{(i)} \ba{c} \mbox{
No thin shell}\\ \ \ \ep_{in}=-1 \ea \\
\hline \mbox{(ii)} \ba{c} \mbox{Thin shell} \\
 \ \xi =0 \ea \ea &\ba{c} \mbox{No thin shell}

 \\ \ep_{in}=+1 \ea \\ \hline
 \ov{\rho}<\rho_{b}&\ba{c} \mbox{Thin shell} \\ \ \
 0<\xi <1\ea&\ba{c} \mbox{Thin shell}\\ \ \ \xi >1 \ea \\
 \hline \ea \]

\newpage\vspace{1cm}
\section*{References}
$[$1$]$ H. Sato, Prog. Theor. Phys., 68, 236 (1982); H. Sato and
K. Maeda, ibid. 70,\\
\hspace*{0.5cm}119 (1983); K. Maeda and H. Sato,
ibid, 70, 772 (1983); 70, 1276 (1983). \hspace*{0.4cm}\\
$[$2$]$ F. Occhionero, L. Veccia-Scavalli, and N. Vittorio, Astron. Astrophys. 97, 169 \\
\hspace*{0.55cm}(1981).\\
$[$3$]$ D. Olson and J. Silk, Astrophys. J., 233, 395 (1979).\\
$[$4$]$ P.J.E. Peebles, Principles of Physical Cosmology
(Princeton University press,\\
\hspace*{0.55cm}Princeton, 1993).\\
$[$5$]$ G. Lemaitre, Ann. Soc. Sci. Bruxelles Ser. I, A53, 51
(1933);R. C. Tolman, \\
\hspace*{0.55cm}Proc. Natl. Acad. Sci.U.S.A. 20, 410 (1934); H. Bondi, Mon. Not. R. Astron.\\
\hspace*{0.55cm}Soc. 107, 343 (1947).\\
$[$6$]$ R. Kantowski, Astrophys. J., 155, 89 (1969).\\
$[$7$]$ H. Sato, Prog. Theor. Phys., 76, 1250 (1986).\\
$[$8$]$ V.A. Berezin, V.A. Kuzmin, and I.I. Tkachev, Phys. Rev. D36, 2919 (1987).\\
$[$9$]$ N. Sakai and K. Maeda, Phys. Rev. D50, 5425 (1994).\\
$[$10$]$ A. Einstein and E. G. Strauss, Rev. Mod. Phys. 17, 120
(1945). \\
$[$11$]$ S. Khakshournia and R. Mansouri(unpublished).
\\

\end{document}